\documentclass[12pt]{article}
\usepackage{amssymb,amsmath,slashed,latexsym}
\usepackage{mathrsfs}
\usepackage{xcolor}
\usepackage{tikz}
\usetikzlibrary{calc}
\usetikzlibrary{matrix}
\usepackage{bbold}
\usepackage{bbm} 
\usepackage[vcentermath]{youngtab}
\usepackage{cite}

\usepackage[T1]{fontenc}

\usepackage{color}
\definecolor{dark-gray}{gray}{0.20}
\definecolor{gray}{gray}{0.30}
\definecolor{light-gray}{gray}{0.80}
\definecolor{dark-red}{rgb}{0.7,0,0}
\definecolor{dark-green}{rgb}{0.1,0.4,0}
\definecolor{dark-blue}{rgb}{0.3,0.3,0.7}
\definecolor{light-blue}{rgb}{0.8,0.8,1}

\pdfoutput=1
\usepackage{hyperref}
\hypersetup{
	colorlinks=true,
	linkcolor=dark-blue,
	citecolor=dark-red,
	urlcolor=dark-green,
	linktoc=page
}

\usepackage{benmacros}

\numberwithin{equation}{section}
\renewcommand{\theequation}{\arabic{section}.\arabic{equation}}

\def\rme{{\rm e}}
\def\rmi{{\rm i}}

\newsavebox{\uuunit}
\sbox{\uuunit}
{\setlength{\unitlength}{0.825em}
	\begin{picture}(0.6,0.7)
	\thinlines
	\put(0,0){\line(1,0){0.5}}
	\put(0.15,0){\line(0,1){0.7}}
	\put(0.35,0){\line(0,1){0.8}}
	\multiput(0.3,0.8)(-0.04,-0.02){12}{\rule{0.5pt}{0.5pt}}
	\end {picture}}

\newcommand{\SU}{\mathop{\rm SU}}
\newcommand{\SO}{\mathop{\rm SO}}
\newcommand{\SL}{\mathop{\rm SL}}
\newcommand{\U}{\mathop{\rm {}U}}
\newcommand{\USp}{\mathop{\rm {}USp}}

\newcommand{\E}{\mathop{\rm {}E}}

\newcommand{\e}{\mathrm{e}}
\newcommand{\w}{\wedge}

\newcommand{\f}[2]{\frac{#1}{#2}}

\usepackage[percent]{overpic}
\usepackage{caption}
\usepackage{subcaption}

\begin{document}  

\begin{titlepage}
 
\bigskip
\bigskip
\bigskip
\begin{center} 
\textbf{\Large  Uplifting GPPZ: A Ten-dimensional Dual of $\mathcal{N}=1^{*}$ }

\bigskip
\bigskip
\bigskip
\bigskip
\bigskip
\bigskip

\textbf{Nikolay Bobev, Fri\dh rik Freyr Gautason, \\[0.2cm]Benjamin E. Niehoff, and Jesse van Muiden }
\bigskip

Instituut voor Theoretische Fysica, KU Leuven,\\ 
Celestijnenlaan 200D, B-3001 Leuven, Belgium
\vskip 5mm

\bigskip
\bigskip
\bigskip
\tt{nikolay.bobev,~ffg,~ben.niehoff,~jesse.vanmuiden~~~@kuleuven.be}  \\
\end{center}

\bigskip
\bigskip
\bigskip
\bigskip

\begin{abstract}

\noindent  
\end{abstract}

We find a new supersymmetric solution of type IIB supergravity which is the uplift of the GPPZ solution of maximal $\SO(6)$ gauged supergravity in five dimensions. This background is expected to be holographically dual to an $\mathcal{N}=1^{*}$ supersymmetric mass deformation of four-dimensional $\mathcal{N}=4$ SYM. The ten-dimensional solution is singular in the region corresponding to the IR regime of the dual gauge theory and we discuss the physics of the singularity in some detail.

\noindent 

\vfill

\end{titlepage}

\newpage

\setcounter{tocdepth}{2}
\tableofcontents

 \section{Introduction}
 \label{sec:intro}
 
Since the early days of holography it was understood that a fruitful way to study the gauge/gravity duality for non-conformal theories is to consider relevant deformations of the four-dimensional $\mathcal{N}=4$ SYM theory, see for example \cite{Girardello:1998pd,Freedman:1999gp}. This approach naturally leads to the study of supergravity solutions which are deformations of the maximally supersymmetric AdS$_5\times S^5$ background of type IIB supergravity. However it turns out to be a difficult task to construct such explicit ten-dimensional solutions, even if one imposes that a certain amount of supersymmetry is preserved. The reason for this is that a general supersymmetric relevant deformation of $\mathcal{N}=4$ SYM breaks part of the $\SO(6)$ R-symmetry of the conformal theory which in turn results in a reduced isometry for the metric and background fields on the internal $S^5$ geometry.

An efficient way to circumvent this technical difficulty is to notice that the lowest-lying Kaluza-Klein (KK) modes in the linearized spectrum of type IIB supergravity around AdS$_5\times S^5$, derived in \cite{Kim:1985ez}, are dual precisely to the most relevant protected operators in the $\mathcal{N}=4$ SYM theory. In addition one can show that these modes are precisely the same AdS$_5$ fields as the ones in the maximal $\mathcal{N}=8$, $\SO(6)$ gauged supergravity of \cite{Gunaydin:1984qu,Gunaydin:1985cu,Pernici:1985ju}. It is then natural to conjecture that the type IIB supergravity equations of motion admit a consistent truncation to precisely these lowest lying KK modes, which is described by the Lagrangian of the five-dimensional gauged supergravity theory. While this was expected to be true for a long time, and supported by preliminary evidence (see for example \cite{Cvetic:2000nc,Pilch:2000ue}), it was only recently that the consistency of this truncation was rigorously established in \cite{Lee:2014mla,Baguet:2015sma}. Moreover, \cite{Hohm:2013vpa,Baguet:2015sma,Baguet:2015xha} give explicit uplift formulae for all ten-dimensional fields in terms of the five-dimensional $\mathcal{N}=8$ supergravity fields, thus allowing for the explicit construction of new type IIB supergravity backgrounds by uplifting any solution of the five-dimensional gauged supergravity.  The feasibility of this approach was recently shown in \cite{Bobev:2018hbq} where a new type IIB supergravity solution dual to the $\mathcal{N}=2^{*}$ SYM theory on $S^4$ was constructed by uplifting the five-dimensional solution found in \cite{Bobev:2013cja}.  In this paper, we shall apply these methods to construct a new family of explicit, ten-dimensional type IIB backgrounds which correspond to the well-known GPPZ flow solutions \cite{Girardello:1999bd} of five-dimensional gauged supergravity, which are holographically dual to an $\mathcal{N}=1^{*}$ deformation of the  $\mathcal{N}=4$ SYM theory.

To construct the ten-dimensional uplift of the GPPZ solutions, we first identify a suitable truncation of the five-dimensional $\mathcal{N}=8$ gauged supergravity in which only a small subset of the 42 scalar fields are kept, and then show that the GPPZ solution is contained within it. A convenient choice for such a truncation is the model with four real scalar fields studied in \cite{Bobev:2016nua}. This truncation is relatively simple and one can easily find explicit expressions for the various matrices that determine the Lagrangian and supersymmetry variations of the five-dimensional theory. This five-dimensional data can then be used in the uplift formulae of \cite{Hohm:2013vpa,Baguet:2015sma,Baguet:2015xha} to arrive at explicit expressions for the metric, axion-dilaton and the NS-NS and R-R form fields of the type IIB supergravity theory. To find relatively compact expressions for all ten-dimensional supergravity fields it is essential to use coordinates on the $S^5$ which are adapted to the symmetry of the problem.  The gauge theory dual of the GPPZ solution is expected to be the particular $\mathcal{N}=1^{*}$ SYM theory obtained from $\mathcal{N}=4$ SYM by deformation of the superpotential by three masses, taken to be real and equal to each other.  Thus the supergravity solution should have an $\SO(3)$ isometry corresponding to the $\SO(3)$ flavor symmetry of the gauge theory.  Choosing coordinates on the $S^5$ adapted to this symmetry, we are then able to write the ten-dimensional uplift of the GPPZ solution in a relatively simple explicit analytic form.  We note that explicit expressions for the metric and axion-dilaton were previously obtained in \cite{Pilch:2000fu}, and the expressions we obtain agree with these.  However, in addition to the metric and axion-dilaton, the full ten-dimensional solution also has non-trivial NS-NS three-form, R-R three-form, and R-R five-form fluxes, which we construct explicitly here.  As a consistency check of our results, we also verify that this background of metric, axion-dilaton, and fluxes, obtained via the uplift formulae of \cite{Hohm:2013vpa,Baguet:2015sma,Baguet:2015xha}, obeys all the equations of motion of type IIB supergravity.

A major motivation for this work is to better understand the beautiful picture of Polchinski and Strassler advanced in \cite{Polchinski:2000uf}.  This paper argues that many supersymmetric vacua of the $\mathcal{N}=1^{*}$ have a holographic description in string theory and supergravity in terms of an asymptotically AdS$_5\times S^5$ solution whose bulk region corresponds to the IR regime of the $\mathcal{N}=1^{*}$ gauge theory, controlled by D3-branes polarized into $(p,q)$ five-branes via the Myers effect \cite{Myers:1999ps}. Naively one may anticipate that the ten-dimensional uplift of the GPPZ solution should be a particular example which realizes the physics advocated by Polchinski and Strassler. However there are problems with this naive expectation. The simplest way to arrive at the puzzle is to note that the five-dimensional GPPZ solution has an integration constant, which we denote by $\lambda$, that can take any real value. Analyzing the asymptotically AdS$_5$ region of the solution, one finds that $\lambda$ is dual to the dimensionless ratio between the gaugino bilinear vev and the mass parameter of the $\mathcal{N}=1^{*}$ theory (with the masses set equal). This appears to be in conflict with the known structure of the space of supersymmetric vacua of the $\mathcal{N}=1^{*}$ theory, which consists of a collection of isolated vacua with discrete values of the gaugino condensate.\footnote{See \cite{Dorey:2000fc,Dorey:1999sj} and in particular \cite{Aharony:2000nt} for a nice discussion relevant for the large $N$ limit of the $\mathcal{N}=1^{*}$ theory.} It is important to note that the structure of the singularity of the GPPZ solution depends on the value of $\lambda$. Thus one possible resolution of the puzzle is that only for a specific set of discrete values of $\lambda$ the GPPZ solution describes a supersymmetric vacuum of the $\mathcal{N}=1^{*}$ theory. Indeed it is possible to argue that only for $|\lambda|\leq 1$ is the singularity of the GPPZ solution physical. This can be done by employing the `Gubser criterion' \cite{Gubser:2000nd} for accepting naked singularities in holographic RG flows of five-dimensional gauged supergravity. Alternatively one can use the `Maldacena-Nu\~nez criterion' \cite{Maldacena:2000mw} for accepting ten-dimensional naked singularities as physical, and apply it to the uplifted ten-dimensional solution we shall construct, to arrive at the same conclusion.  However, the puzzle of having the continuous parameter $\lambda$ in the range $0 \le \lambda \leq 1$ remains, and one hopes that the ten-dimensional uplift of the GPPZ solutions might resolve it.  As already noted in \cite{Pilch:2000fu}, for $0\leq \lambda<1$ the IR singularity in the ten-dimensional metric is milder than the one in five dimensions --- rather than being singular over the whole $S^5$ at a constant value of the holographic coordinate, the solution is singular only over a ring-like equator.  The behavior of all of the supergravity fields near this singularity are inconsistent with the suggested interpretation in \cite{Pilch:2000fu} of D7-branes, but it remains difficult to see the expected $(p,q)$ 5-branes described in \cite{Polchinski:2000uf}. Thus the interpretation of the solutions with $0\leq \lambda <1$ as holographic duals to supersymmetric vacua  of the $\mathcal{N}=1^{*}$ theory remains unclear.  These results suggest that perhaps only for $\lambda=1$ can we hope to interpret the singular uplifted GPPZ solution as a supergravity background dual to a supersymmetric vacuum of the $\mathcal{N}=1^{*}$ theory. While this has been suggested before in \cite{Gubser:2000nd}, \cite{Pilch:2000fu} and \cite{Aharony:2000nt} using different arguments, the analysis of the singular region of our full ten-dimensional solution for $\lambda=1$ is subtle and we are unfortunately unable to find a fully satisfactory resolution of the problem. Nevertheless we believe that our results provide a stepping stone towards explicit supergravity solutions that realize the Polchinski-Strassler scenario.  

We continue in the next section with a summary of the consistent truncation of five-dimensional $\mathcal{N}=8$ gauged supergravity and present the GPPZ solution of \cite{Girardello:1999bd}. In Section~\ref{sec:10d} we present the explicit uplift of this solution to ten-dimensional type IIB supergravity and discuss in some detail the singularity in the IR region. We conclude in Section~\ref{sec:discussion} with a short discussion on several problems for further research.

\bigskip 
 
 \textit{Note added:} While we were completing this manuscript the paper \cite{Petrini:2018pjk}, which has overlap with our results, appeared on the arXiv. We have subsequently communicated with the authors of \cite{Petrini:2018pjk} and have verified that our type IIB supergravity solution agrees with theirs upon a change of coordinates. We are grateful to the authors of \cite{Petrini:2018pjk} for the useful discussions about the comparison of the two solutions.

 \section{The five-dimensional solution}
 \label{sec:5d}

It is well-known that the $\mathcal{N}=4$ SYM theory admits superpotential deformations that break conformal invariance while preserving $\mathcal{N}=1$ supersymmetry, see for example \cite{Leigh:1995ep}. To write the Lagrangian for this family of supersymmetric gauge theories it is convenient to organize the fields of the $\mathcal{N}=4$ theory into $\mathcal{N}=1$ multiplets. One has a vector multiplet, $\mathcal{V}^{a}$ with gauge field $A_{\mu}^{a}$ and a gaugino $\lambda^a$, together with  three chiral superfields $\Phi_{i}^a$ each of which contains a complex scalar field $Z_i^a$ and a fermion $\chi_i^a$. The index $a$ is in the adjoint of the gauge group\footnote{In this paper we choose the gauge group to be $\SU(N)$.} and $i=1,2,3$. In addition to the usual kinetic terms for the chiral multiplets and the vector multiplet one can add the following superpotential\footnote{We use the notation and conventions of \cite{Bobev:2016nua} and refer the reader to that paper for more details on the supersymmetric Lagrangian of this theory.}
\begin{equation}\label{eq:suppotN1*}
W = \sqrt{2}g_{\rm YM}f^{abc}\Phi_1^{a}\Phi_2^{b}\Phi_3^{c} + \frac{1}{2}\sum_{i=1}^{3}m_i\Phi_i^a\Phi_i^a\,.
\end{equation}
Here $g_{\rm YM}$ is the Yang-Mills coupling, $f^{abc}$ are the structure constants of the gauge group and summation over repeated indices is assumed. 
 
This mass deformation of $\mathcal{N}=4$ SYM is sometimes referred to as $\mathcal{N}=1^{*}$ although this nomenclature is imperfect since we actually have three independent complex mass parameters in the superpotential in \eqref{eq:suppotN1*} and thus a three-parameter family of theories. This theory has a rich space of discrete supersymmetric vacua which has been studied in many references, see for example  \cite{Dorey:1999sj,Dorey:2000fc,Polchinski:2000uf,Aharony:2000nt}. In the limit $m_i=0$ the superpotential in \eqref{eq:suppotN1*} reduces to the one of $\mathcal{N}=4$ SYM and in this $\mathcal{N}=1$ formulation of the theory only an $\SU(3)\times\U(1)$ subgroup of the $\SO(6)$ R-symmetry of the theory is manifest. For general values of the masses $m_i$ the $\SU(3)$ symmetry is broken explicitly and in a general supersymmetric vacuum of the theory the $\U(1)$ symmetry is spontaneously broken. Our interest in this work is in the limit of the model above where all three masses are taken to be real, $m_i=\bar{m}_i$, and equal, $m_1=m_2=m_3$. In this case an $\SO(3)$ subgroup of the $\SU(3)$ discussed above is preserved. We now proceed to summarize this symmetry enhancement as it is valuable when one studies the supergravity dual of this gauge theory.
 
 \subsection{The consistent truncation}
 \label{subsec:truncation}

The holographic dual of  $\mathcal{N}=4$ SYM  in the planar limit and at large 't Hooft coupling is given by the maximally supersymmetric AdS$_5\times S^5$ solution of type IIB supergravity. It is expected that the dual of the supersymmetric deformation of $\mathcal{N}=4$ SYM in \eqref{eq:suppotN1*} is given by a deformation of this solution which preserves 1/4 of the maximal sypersymmetry and breaks part of the isometry of AdS$_5\times S^5$. Constructing this supergravity solution directly in ten dimensions is a formidable task since the type IIB supergravity BPS equations are a system of non-linear PDEs. A useful strategy to make progress is to use the fact that the lowest lying Kaluza-Klein modes of the supergravity spectrum around AdS$_5\times S^5$ admit an effective five-dimensional description in terms of the maximal $\SO(6)$ gauged supergravity found in \cite{Gunaydin:1984qu,Gunaydin:1985cu,Pernici:1985ju}. This consistent truncation of the equations of motion of type IIB supergravity was recently established rigorously in \cite{Lee:2014mla,Baguet:2015sma}. To this end it is useful to summarize some of the salient features of the five-dimensional supergravity theory of \cite{Gunaydin:1984qu,Gunaydin:1985cu,Pernici:1985ju}.

The bosonic fields of the five-dimensional supergravity are the metric, 12 two-form potentials $B_{\mu\nu}$, the $\SO(6)$ gauge field $A_{\mu}$, as well as 42 scalar fields. We are interested in solutions of the theory for which the fields $B_{\mu\nu}$ and $A_{\mu}$ vanish so we ignore these fields from now on. The theory is invariant under the maximal subgroup, $\SL(6,\mathbb{R})\times \SL(2,\mathbb{R})$, of the  $\E_{6(6)}$ symmetry of the ungauged supergravity and all matter fields in the theory are in various representations of this maximal subgroup. The 42 scalars of interest here live in the coset $\E_{6(6)}/\USp(8)$. Following \cite{Gunaydin:1985cu} one can parametrize this coset space by the $27\times 27$ matrix
 \begin{align}\label{eq:Xhatmatrix}
 \hat{X} = \left( \begin{array}{cc}
 -4 \Lambda_{[I}^{\phantom{[I}[P}\delta_{J]}^{Q]}  & \sqrt{2} \Sigma_{IJR\beta} \\ 
 \sqrt{2}\Sigma^{PQK\alpha} & \Lambda_R^{\phantom{[I}K}\delta_{\beta}^{\alpha} + \Lambda_{\beta}^{\phantom{\beta}\alpha} \delta^{K}_{R}
 \end{array}  \right).
 \end{align}
The capital Latin indices above transform under $\SL(6,\mathbb{R})$ and the lower-case Greek indices transform under $\SL(2,\mathbb{R})$. The matrix $\Lambda_{I}^{\phantom{I}J}$ is $6\times 6$ symmetric and traceless, $\Lambda_{\alpha}^{\phantom{\alpha}\beta}$ is $2\times 2$ symmetric and traceless, while the tensor $\Sigma_{IJK\alpha}$ has $20$ independent components since it is completely antisymmetric and self-dual in the indices $IJK$ for any $\alpha=1,2$.

The action and supersymmetry variations of the five-dimensional theory are written in terms of the vielbein on the scalar manifold defined by exponentiating the matrix in \eqref{eq:Xhatmatrix}, $U = \rme^{\hat{X}}$. To apply the uplift formulae of \cite{Hohm:2013vpa,Baguet:2015sma,Baguet:2015xha} it proves useful to work directly with the metric on the $\E_{6(6)}/\USp(8)$ coset space defined by
 \begin{align}\label{eq:mdef}
 M = U\cdot U^T~.
 \end{align}
The $27\times27$ components of $M$ can be organized into representations of $\SL(6,\mathbb{R})\times \SL(2,\mathbb{R})$ as follows
\begin{align}\label{eq:scalarmatrix}
M=
\left(\begin{matrix}
M_{IJ,PQ} & M_{IJ}^{\phantom{IJ}R\beta} \\ 
M^{K \alpha}_{\phantom{K\alpha}PQ} & M^{K \alpha, R \beta}
\end{matrix}\right)\,.
\end{align}
In the expression above the capital Latin index pairs $IJ$ and $PQ$ are antisymmetric and thus transform in the ${\bf 15}$ of $\SL(6,\mathbb{R})$. The index pair consisting of a Latin and a Greek index is in the $({\bf 6},{\bf 2})$ representation of $\SL(6,\mathbb{R})\times \SL(2,\mathbb{R})$. The explicit knowledge of the components of the matrix $M$ in \eqref{eq:scalarmatrix} is important when one wants to utilize the uplift formulae of \cite{Baguet:2015sma}.

Working with all 42 scalars of the gauged supergravity theory is a daunting task therefore one has to look for a suitable truncation of the five-dimensional theory to a more manageable set of fields. Fortunately the gauge theory provides the necessary hints to get oriented in this problem. The $\mathcal{N}=1^{*}$ theory superpotential in \ref{eq:suppotN1*} can be expanded in components and then one finds that there are 3 bosonic bilinears, 6 real fermionic bilinears as well as the complexified gauge coupling appearing in the full Lagrangian of the gauge theory. In addition one should recall that the complex gaugino bilinear vev can be dynamically generated along the RG flow to the IR and is thus in general non-zero in a supersymmetric vacuum of the $\mathcal{N}=1^{*}$ theory.  This simple gauge theory analysis implies that we should expect that the five-dimensional supergravity dual of the most general $\mathcal{N}=1^{*}$ deformation in \eqref{eq:suppotN1*} should have at least 13 real scalar fields. There is however a small subtlety since the bosonic bilinear operator $|Z_1|^2+|Z_2|^2+|Z_3|^2$, known as the Konishi operator, is not protected and is expected to be dual to a stringy rather than a supergravity KK mode. Thus we conclude that we should need at least 12 real supergravity scalar fields to study the holographic dual of the $\mathcal{N}=1^{*}$ theory. In \cite{Bobev:2016nua} a truncation of the gauged supergravity theory was discussed which contains 18 real scalar fields and contains precisely the 12 scalars discussed above, see Appendix B of \cite{Bobev:2016nua}. This truncation was obtained following the original idea of \cite{Warner:1983vz} to look at the subsector of the gauged supergravity theory invariant under a subgroup of the $\SO(6)\times\SL(2,\mathbf{R})$ symmetry. The slight subtlety is that to obtain the 18-scalar truncation in \cite{Bobev:2016nua} one has to use discrete rather than continuous subgroups of $\SO(6)\times\SL(2,\mathbf{R})$. At this point we should remember that our goal here is to study a limit of the $\mathcal{N}=1^{*}$ theory  in which the three masses in \eqref{eq:suppotN1*} are equal and real. This results in a drastic simplification of the problem and it turns out to be sufficient to work with a further four-scalar truncation of the 18-scalar model in \cite{Bobev:2016nua}.\footnote{In \cite{Pilch:2000fu} two consistent truncations of the five-dimensional supergravity were used to obtain the GPPZ solution. One of these truncations agrees with the four-scalar model considered here although the arguments used to arrive at it are slightly different than in \cite{Bobev:2016nua}. Of course the form of the solution is independent of the consistent truncation used to construct it.} To obtain this truncation one first has to use an additional discrete symmetry which effectively implements the reality condition on the masses in  \eqref{eq:suppotN1*} and then impose an $\SO(3)$ symmetry, see \cite{Bobev:2016nua} for further details.

We will use the notation adopted in \cite{Bobev:2016nua} and present some of the details on the construction of this four-scalar truncation of the $\mathcal{N}=8$ gauged supergravity. The matrix $\Lambda_{J}^{\phantom{J}I}$ used in the scalar coset generator \eqref{eq:Xhatmatrix} transforms in the $\mathbf{20}'$ of $\SO(6)$ and in our truncation takes the simple diagonal form
\begin{equation}
\Lambda_{J}^{\phantom{J}I} = \text{diag}(\bar{\alpha}_1,-\bar{\alpha}_1,\bar{\alpha}_1,-\bar{\alpha}_1,\bar{\alpha}_1,-\bar{\alpha}_1)\,,
\end{equation}
where $\bar{\alpha}_1$ is a real scalar. The matrix $\Lambda_{\alpha}^{\phantom{\alpha}\beta}$ in \eqref{eq:Xhatmatrix} transforms only under $\SL(2,\mathbf{R})$ and for the truncation at hand reads
\begin{equation}
\Lambda_{1}^{\phantom{1}1}=-\Lambda_{2}^{\phantom{2}2}=\bar{\varphi}\,,
\end{equation}
with $\bar{\varphi}$ a real scalar field. The non-vanishing components of the tensor $\Sigma_{IJK\alpha}$ in the $\mathbf{10}\oplus\overline{\mathbf{10}}$ representation of $\SO(6)$ are
\begin{equation}
\begin{split}
\Sigma_{1351} &= -\Sigma_{2462}=\frac{1}{2}(3\bar{\phi}_1-\bar{\phi}_4)\,,\\
\Sigma_{1461}&=\Sigma_{2361}=\Sigma_{2451} = -\Sigma_{2352}= -\Sigma_{1452}= -\Sigma_{1362} = \frac{1}{2}(\bar{\phi}_1+\bar{\phi}_4)\;,
\end{split}
\end{equation}
where we have introduced two other real scalar fields, $\bar{\phi}_1$ and $\bar{\phi}_4$.

Around the maximally supersymmetric AdS$_5$ vacuum of the gauged supergravity the scalar fields defined above are dual to gauge invariant protected operators in the $\mathcal{N}=4$ theory. The scalar $\bar{\varphi}$ is dual to an operator of conformal dimension $4$ which is simply the Yang-Mills kinetic term in the SYM theory. The scalar $\bar{\alpha}_1$ is dual to the dimension 2 bosonic bilinear operator $\sum_{i=1}^{3}\text{Tr}(Z_i^2+\bar{Z}_i^2)$. The scalar $\bar{\phi}_1$ is dual to  the fermionic bilinear operator $\sum_{i=1}^{3}\text{Tr}(\chi_i\chi_i+\bar{\chi}_i\bar{\chi}_i)$ which has dimension 3. Finally the scalar $\bar{\phi}_4$ is the supergravity dual of the gaugino bilinear operator $\text{Tr}(\lambda\lambda+\bar{\lambda}\bar{\lambda})$ which also has dimension 3. To describe the five-dimensional GPPZ solution in \cite{Girardello:1999bd} we only need the scalars $\bar{\phi}_1$ and $\bar{\phi}_4$. In the next section we explain in some detail how to explicitly obtain this solution in the four-scalar model described above. Before that however we note that it proves convenient to introduce new scalar variables which significantly simplify the supergravity Lagrangian in the truncated model. To this end we define the combinations
\begin{equation}\label{eq:scalarredef}
\bar{\varphi} +\rmi \bar{\phi}_4 = \frac{1}{4}(r_1\e^{\rmi \zeta_1} - 3 r_2\e^{\rmi \zeta_2})\,, \qquad \bar{\alpha}_1 +{\rm i} \bar{\phi}_1 = \frac{1}{4}(r_1\e^{\rmi \zeta_1} + r_2\e^{\rmi \zeta_2}) \,,
\end{equation}
and then introduce the complex scalars $z_i$ which parametrize two copies of the Poincar\'e disc
\begin{equation}
z_j = \tanh(r_j/2)\e^{\rmi \zeta_j}\,.
\end{equation}
We now proceed to present explicitly the Lagrangian and supersymmetry variations of this four-scalar truncation of the $\mathcal{N}=8$ gauged supergravity.

 \subsection{The GPPZ solution}
 \label{subsec:gppz}

The Lagrangian of our four-scalar truncation is\footnote{Following the conventions in \cite{Gunaydin:1985cu} we work in mostly minus signature.}
\begin{equation}\label{eq:Lag4scalar}
{\cal L} =\sqrt{|g|}\left( -\frac{1}{4} R -\f12 {\cal K}_{i\bar{\jmath}}~ \partial_{\mu} z^i  \partial^{\mu} \bar{z}^{\bar \jmath} - {\cal P}\right)~,
\end{equation}
where the K\"ahler potential and metric on the $[\SU(1,1)/\U(1)]^2$ scalar manifold are given by
\begin{equation}
{\cal K}_{i\bar{\jmath}} = \partial_i \partial_{\bar \jmath} {\cal K}~,\qquad {\cal K} = -\log(1-z_1\bar{z}_1)(1-z_2\bar{z}_2)^3~.
\end{equation}
The potential of this model can be written in terms of a holomorphic superpotential $\mathcal{W}$ in the standard way
\begin{equation}
{\cal P} = \f12 \e^{\cal K}\left[{\cal K}^{i\bar{\jmath}}D_i{\cal W}D_{\bar \jmath} \overline{\cal W}-\f83 {\cal W}\overline{\cal W}\right]~,\qquad {\cal W} = \f{3g}{4}(1+z_1z_2)(1-z_2^2)~,
\end{equation}
where $D_i$ denotes the K{\"a}hler covariant derivative defined by $D_i(\cdot) = (\partial_{i} +\partial_{i}\mathcal{K})(\cdot)$. It is easy to show that there are two AdS$_5$ vacua in this truncation by studying the critical points of the potential. One of the vacua is the maximally supersymmetric $\SO(6)$ invariant vacuum dual to the conformal vacuum of $\mathcal{N}=4$ SYM
\begin{equation}\label{eq:AdS5SO6}
z_1=z_2=0\,,\qquad \mathcal{P} = -\frac{3g^2}{4}\,. 
\end{equation}
The other vacuum is not supersymmetric and preserves an $\SU(3)$ subgroup of $\SO(6)$, see for example \cite{Khavaev:1998fb},
\begin{equation}\label{eq:AdS5SU3}
z_1=  - z_2= \rmi \tanh\left(\frac{\log(2\pm \sqrt{3})}{4}\right)\,,\qquad \mathcal{P} = -\frac{27g^2}{32}\,. 
\end{equation}
This vacuum is perturbatively unstable since some of the 42 scalars of the five-dimensional supergravity theory have masses which are below the BF bound \cite{KP}. 

We are interested in supersymmetric domain wall solutions of the truncation described above. To this end we make an ansatz for the metric of the form
\begin{equation}
\dd s_5^2 = -\dd r^2 + \e^{2A}\dd s_4^2~,
\end{equation}
where $\dd s_4^2$ denotes the Minkowski metric and we assume that the scalars depend only on the radial coordinate $r$. In these coordinates an AdS$_5$ vacuum is given by setting the scalars to their critical values in \eqref{eq:AdS5SO6} or \eqref{eq:AdS5SU3} and taking $A = r/L$. Here $L$ is the length scale of AdS$_5$ which is determined by the value of the potential at the critical point in \eqref{eq:AdS5SO6} or \eqref{eq:AdS5SU3} through the relation $L^2 = -3/\mathcal{P}$.

With this Ansatz at hand one can analyze the supersymmetry variations of the five-dimensional maximal gauged supergravity \cite{Gunaydin:1985cu} and derive the following set of BPS equations
\begin{align}
(A')^2 &= \f{4}{9} \e^{\cal K}  {\cal W}\overline{\cal W}~,\label{ABPS}\\
(A')(z^i)' &= -\f{2}{3}\e^{\cal K} {\cal W}{\cal K}^{i\bar{\jmath}}D_{\bar{\jmath}}\overline{\cal W}~,\label{zBPS}\\
(A')(\bar{z}^{\bar{\imath}})' &= -\f{2}{3}\e^{\cal K} \overline{\cal W}{\cal K}^{\bar{\imath}j}D_{j}{\cal W}~.\label{zbBPS}
\end{align}
One can then show that these ODEs imply that all equations of motion are satisfied. This is perhaps most easily shown by noting that one can write the Lagrangian in \eqref{eq:Lag4scalar}, up to boundary terms, as a sum of the BPS equations squared
\begin{align}
{\cal L} &= \sqrt{|g|}\left[- \f12 {\cal K}_{i\bar{\jmath}}\left((z^i)' + 2\e^{{\cal K}/2} \sqrt{\cal W}{\cal K}^{i\bar{m}}D_{\bar{m}}\sqrt{\overline{\cal W}}\right)\left((\bar{z}^{\bar{\jmath}})' +2 \e^{{\cal K}/2} \sqrt{\overline{\cal W}}{\cal K}^{\bar{\jmath}n}D_{n}\sqrt{\cal W}\right)\right.\nonumber\\
&\left.\qquad+3\left(A' - \f{2}{3} \e^{{\cal K}/2}  \sqrt{{\cal W}\overline{\cal W}}\right)^2 \right] - \partial_r\left(\sqrt{|g|}\e^{{\cal K}/2}  \sqrt{{\cal W}\overline{\cal W}}\right)~.
\end{align}
In writing this we have made a choice of sign when we take the square root of the equation in \eqref{ABPS} such that the $\SO(6)$ invariant AdS$_5$ vacuum in \eqref{eq:AdS5SO6} is located at large positive values of the radial coordinate $r$.

Equipped with the explicit BPS equations it is possible to show that they admit a consistent truncation in which  we set $z_i=-\bar{z}_i$, or equivalently $\zeta_1=\zeta_2=\pi/2$ (c.f.  \eqref{eq:scalarredef}). This truncated model with two scalars is precisely the one used in \cite{Girardello:1999bd} to construct the five-dimensional solution dual to the equal mass $\mathcal{N}=1^{*}$ SYM theory in \eqref{eq:suppotN1*}. An alternative way to arrive at this two-scalar truncation of the five-dimensional $\mathcal{N}=8$ supergravity by using discrete symmetries was described in \cite{Pilch:2000fu}.

To describe the solution of the BPS equations for this two-scalar model we find it convenient to rewrite the scalars in terms of new functions as
\begin{equation}
z_1 =\rmi\f{G^5+F^3}{G^5-F^3}~,\qquad z_2 = \rmi\f{G+F}{G-F}~.
\end{equation}
In terms of $G$ and $F$ the BPS equations in \eqref{ABPS}-\eqref{zbBPS} take the form
\begin{align}
A' &= -\f{g}8 \f{(G^2+F^2)(G^4+F^2)}{F^2G^3}~,\\
G' &= \f{3 g}{4}(G^2-1)~,\\
(F^2)' &= \f{g}{4 G^3}\left(F^2 - G^6 + 9F^2G^2(G^2-1)\right)~.
\end{align}
These equations can be solved by
\begin{align}
\e^{2A} &= \left(\f{1}{t}-t\right)\left(\f{1}{t^3}-\lambda^2 t^3\right)^{1/3}~,\\
G &= \f{1-\lambda t^3}{1+\lambda t^3}~,\\
F^2 &= G^3\f{1-t}{1+t}~,
\end{align}
where $\lambda$ is an integration constant and we have introduced a new radial variable defined through the equation $2t' =g t$.\footnote{We choose to use the new radial variable $t$ to comply with the notation used in \cite{Pilch:2000fu}.} In the coordinate  $t$ the metric takes a simple form
\begin{equation}\label{sugrametric}
\dd s_5^2 = \f{4}{g^2t^2}\left(- \dd t^2+ \left(1-t^2\right)\left(1-\lambda^2 t^6\right)^{1/3}\dd s_4^2\right)\,.
\end{equation}
For small values of $t$ the solution approaches the maximally supersymmetric AdS$_5$ vacuum in \eqref{eq:AdS5SO6} with length scale $L = \f{2}{g}$.

To compare with the results presented in \cite{Pilch:2000fu} it is useful to define yet another parametrization of the two scalar fields in our model. It is obtained by setting $G=\mu^{-2}$ and $F= \mu^{-3} \nu^{-1}$ which in turn implies
\begin{equation}
z_1 =\rmi~\f{\mu-\nu^3}{\mu+\nu^3}~,\qquad z_2 = \rmi~\f{1-\mu\nu}{1+\mu\nu}~.
\end{equation}
One of the benefits of this new parametrization of the scalars is that the asymptotic expansion near the UV AdS$_5$ vacuum, i.e. for small $t$, takes the simple form
\begin{equation}
\mu \approx 1+ \lambda t^3~, \qquad \nu \approx 1+t~.
\end{equation}
From this expansion we note that $\nu$ is the scalar degree of freedom dual to an operator of conformal dimension $\Delta=3$ which has a non-trivial source, i.e. the fermionic bilinear operator added to the Lagrangian by the superpotential deformation in \eqref{eq:Lag4scalar}. By the same token $\mu$ is also dual to an operator with the same conformal dimension however this operator has only a non-trivial vev controlled by the constant $\lambda$. This is simply the vev of the gaugino bilinear in the $\mathcal{N}=1^{*}$ theory.

The integration constant $\lambda$ in the supergravity solution should then be dual to the dimensionless ratio of the gaugino bilinear vev and the mass parameter in the superpotential \eqref{eq:Lag4scalar}. Given that the $\mathcal{N}=1^{*}$ theory has a collection of isolated supersymmetric vacua, i.e. there is no moduli space, it is natural to expect that not all values of the parameter $\lambda$ are physically acceptable. To understand this better we can choose the sign of the radial coordinate $t$ to be positive and find that the coordinate range of $t$ depends on the value of $\lambda$ as follows
\begin{equation}\label{eq:trange}
\begin{array}{cl}
t\in (0,1] & \text{for }|\lambda| \le 1~,\\
t\in (0,|\lambda|^{-1/3}]&\text{for }|\lambda|>1~.
\end{array}
\end{equation}
The five-dimensional metric has a naked curvature singularity when $t$ approaches the upper end of the coordinate range. It was advocated in \cite{Gubser:2000nd} that some singularities in asymptotically AdS supergravity solutions should be accepted as physical since they can describe the IR physics of a gapped or free gauge theory. The criterion for an acceptable singularity proposed in \cite{Gubser:2000nd} is to evaluate the potential $\mathcal{P}$ of the five-dimensional supergravity Lagrangian in \eqref{eq:Lag4scalar} and impose that it is bounded from above. Applying this criterion to the explicit solution at hand one finds that $\lambda$ has to obey
\begin{equation}\label{eq:lambdagood}
|\lambda| \le 1~.
\end{equation}
The same conclusion for the range of $\lambda$ leading to an acceptable singularity was reached in \cite{Girardello:1999bd}. Given that the $\mathcal{N}=1^{*}$ theory has only isolated vacua it is still puzzling to find a continuous range of acceptable values for the parameter $\lambda$. Indeed there are several hints in the literature, both from the point of view of the gauge theory \cite{Aharony:2000nt} as well as from supergravity \cite{Gubser:2000nd,Pilch:2000fu}, that only the value $\lambda=1$ in the range \eqref{eq:lambdagood} corresponds to a supergravity solution dual to a supersymmetric vacuum of the equal-mass $\mathcal{N}=1^{*}$ theory. In the next section we uplift this five-dimensional supergravity solution to ten-dimensional type IIB supergravity where some further features of the IR singularity will be manifest.

\section{The ten-dimensional solution}
\label{sec:10d}

In this section we uplift the five-dimensional GPPZ solution presented in Section~\ref{subsec:gppz} using the uplift formulae derived in \cite{Baguet:2015sma}. We start our discussion with a rapid review of how the uplift procedure of \cite{Baguet:2015sma} works given the explicit $E_{6(6)}$ matrices of the five-dimensional gauged supergravity theory presented in Section \ref{subsec:truncation}. Our conventions for type IIB supergravity are the same as the ones summarized in Appendix C of \cite{Bobev:2018hbq}. Note also that in this section we work with mostly plus signature of the ten-dimensional metric. In particular this implies that when we use the five-dimensional metric from Section~\ref{sec:5d}, see \eqref{sugrametric}, we first have to change the signature of the five-dimensional metric by hand.

The ten-dimensional metric is a warped product of the five-dimensional supergravity metric in \eqref{sugrametric} and a squashed metric on the five-sphere
\begin{equation}\label{eq:fullmetricuplift}
\dd s_{10}^2 =\Delta^{-2/3}\left(\dd s_5^2 + \dd \Omega_5^2\right)~,
\end{equation}
where the warp-factor $\Delta$ will be defined shortly. The (inverse) metric on $S^5$ is given directly in terms of the five-dimensional scalar matrix $M$ in \eqref{eq:scalarmatrix} as follows
\begin{equation}\label{eq:internalmetricuplift}
G^{mn} = {\cal K}_{IJ}^{\phantom{IJ}m}{\cal K}_{PQ}^{\phantom{PQ}n} M^{IJ,PQ}~.
\end{equation}
In this equation we have introduced the Killing-vectors on $S^5$ defined by 
\begin{equation}\label{eq:Killinvecdef}
{\cal K}_{IJ}^{\phantom{IJ}m}=-\frac{g}{2}\widehat{G}^{mn}Y_{[I}\nabla_n Y_{J]}~,
\end{equation}
where the coordinates $Y_I$ for $I=1,\dots,6$ define an embedding of $S^5$ into $\mathbf{R}^6$ and $\widehat{G}^{mn}$ is the inverse round metric on $S^5$ with unit radius. Explicitly we use coordinates on $S^5$ that make the breaking $\SU(4)\to \SO(3)\times \U(1)$ manifest:
\begin{align}\label{eq:theYs}
 \left(\begin{matrix}
Y_1 + \rmi Y_{2} \\ 
Y_3 + \rmi Y_{4} \\ 
Y_5 + \rmi Y_{6}
\end{matrix}\right)=  \rme ^{ \rmi \alpha} \cos \chi  ~{\cal R} \left(\begin{matrix}
1 \\ 
0 \\ 
0
\end{matrix}\right) + \rmi ~\rme ^{ \rmi \alpha}\sin \chi  ~{\cal R} \left(\begin{matrix}
0 \\ 
1 \\ 
0
\end{matrix}\right),
\end{align}
where ${\cal R}=\rme^{\omega g_1} \rme^{\xi_1 g_2} \rme^{\xi_2 g_1}$ is an $\SO(3)$ rotation matrix, parametrized in terms of the Euler angles $\left( \omega, \xi_1, \xi_2 \right)$ and the $\SO(3)$ generators $g_1,g_2$:
\begin{align}
g_1 = \left(\begin{matrix}
0 & -1 & 0 \\ 
1 & 0 & 0 \\ 
0 & 0 & 0
\end{matrix}\right), \qquad g_2 = \left(\begin{matrix}
0 & 0 & 1 \\ 
0 & 0 & 0 \\ 
-1 & 0 & 0
\end{matrix}\right).
\end{align}
In these coordinates the round metric on $S^5$ is written as a $\U(1)$-bundle over $\mathbb{C}P^2$
\begin{align}
\dd \widehat \Omega_5^2 &= \dd s_{\mathbb{C}P^2}^2 + (\dd\alpha+\sigma_3 \sin 2\chi )^2\,,\label{eq:roundmetric}\\
\dd s_{\mathbb{C}P^2}^2 &= \dd\chi^2+\sin ^2\chi~\sigma_1^2  + \cos ^2\chi~\sigma_2^2  +\cos ^22\chi~\sigma_3^2 ~,
\end{align}
where the $\SO(3)$ left-invariant one-forms $\sigma_i$ are explicitly given by
\begin{equation}
\begin{split}
\sigma_1 =&  \cos \xi_2\sin\xi_1\,\dd \omega - \sin \xi_2\, \dd\xi_1~, \\
\sigma_2 =&  -\sin \xi_2\sin \xi_1\, \dd\omega-\cos\xi_2\,\dd\xi_1~,\\
\sigma_3 =& -\cos \xi_1 \,\dd \omega - \dd \xi_2~,
\end{split}
\end{equation}
and satisfy 
\begin{equation}
\dd \sigma_i = \frac12 \varepsilon_{ijk} \sigma_j \w \sigma_k~.
\end{equation}
The axion and dilaton are packaged together in an $\SU(1,1)/\U(1)$ matrix $m_{\alpha\beta}$
\begin{equation}\label{eq:dilatonaxiondef}
m_{\alpha\beta} =\begin{bmatrix}\e^{\Phi} (C_0)^2 + \e^{-\Phi} & -\e^{\Phi}C_0\\
-\e^\Phi C_0& \e^{\Phi}\end{bmatrix}~,
\end{equation}
which has determinant one by construction. The inverse of this matrix is obtained from the matrix $M$ of the five-dimensional theory in a similar way as the metric
\begin{equation}\label{eq:dilatonaxionuplift}
m^{\alpha\beta} = \Delta^{4/3}Y_IY_J M^{I\alpha,J\beta}~.
\end{equation}
This equation also implicitly defines the metric function $\Delta$ that can be obtained by setting the determinant of \eqref{eq:dilatonaxionuplift} to one. The type IIB two-forms and four-forms have similar compact expressions in terms of the five-dimensional data
\begin{align}
A_{mn}^{\phantom{mn}\alpha} &= -\frac{2}{g}\varepsilon^{\alpha\beta} G_{nk} {\cal K}_{IJ}^{\phantom{IJ}k}M^{IJ}_{\phantom{IJ}P\beta}\partial_m Y^P~,\label{eq:twoformuplift}\\
C_{klmn} &= \frac{4}{g^4}\left(\sqrt{\widehat{G}}~\varepsilon_{klmnp}\widehat{G}^{pq}\Delta^{4/3}m_{\alpha\beta}\partial_q(\Delta^{-4/3}m^{\alpha\beta})+\hat{\omega}_{klmn}\right)~,
\end{align}
where the four-form $\hat{\omega}$ is defined in terms of the volume form on the rounds $S^5$ via
\begin{equation}
\dd\hat\omega = 16\text{vol}_{S^5} = \dd\left(\cos 4\chi~\sigma_1\w\sigma_2\w\sigma_3\w\dd\alpha\right)~.
\end{equation}
From \eqref{eq:twoformuplift} we extract the NS-NS 2-form $B_2 = A^1$ and the R-R 2-form $C_2 = A^2$.

\subsection{Explicit uplift of GPPZ}
\label{sec:gppzuplift}

Combining the supergravity data in Section \ref{sec:5d} with the uplift formulae reviewed above and the coordinates defined in \eqref{eq:theYs} provides a full solution of type IIB supergravity. It is a non-trivial exercise to write this ten-dimensional background in a compact form. To facilitate this it is convenient to define the four functions
\begin{equation}\label{eq:Kdef}
\begin{split}
K_1 &= (1+t^2)(1-\lambda^2 t^8)+ 2t^2\left((1-\lambda^2 t^6)- \lambda t^2(1-t^2)\cos(4\alpha)\right)\cos2\chi~,\\
K_2 &= (1+t^2)(1-\lambda^2 t^8)- 2t^2\left((1-\lambda^2 t^6)- \lambda t^2(1-t^2)\cos(4\alpha)\right)\cos2\chi~,\\
K_3 &= 2\lambda t^4(1-t^2)\cos2\chi~\sin4\alpha~,\\
K_4 &= (1+t^2)^2(1+\lambda t^4)^2  -4t^4(1+\lambda t^2)^2\cos^2 2\chi~.
\end{split}
\end{equation}
Using these functions the uplifted Einstein frame metric can be written in a relatively compact form\footnote{In all supergravity fields in this section we have introduced explicit factors of the string coupling $g_s$ to comply with the conventions used in \cite{Bobev:2018hbq}.}
\begin{equation}\label{eq:10dmetexpl}
\dd s_{10}^2 = \frac{(K_1 K_2 - K_3^2)^{1/4}}{\sqrt{ g_s }}\left(\frac{\dd s_5^2}{(1-t^2)\sqrt{(1-\lambda^2t^6)}}+ \frac{4\sqrt{1-\lambda^2 t^6}}{g^2(K_1 K_2 -K_3^2)}~\dd \Omega^2_5\right)\,,
\end{equation}
where $\dd s_5^2$ is given in \eqref{sugrametric} and the squashed metric on $S^5$ can be written as
\begin{align}\label{eq:sqspheremetexpl}
\dd \Omega^2_5 =&~K_4 \dd\chi^2 - 4\lambda t^4(1-t^2)^2(\cos2\alpha~\dd\chi-\sin2\alpha~\cos2\chi~\sigma_3)^2\nonumber\\
&- 4\lambda t^6~\dd (\cos2\alpha~\cos2\chi)^2  +\frac{(1-\lambda^2 t^8)^2(1-t^2)}{(1-\lambda^2t^6)}(\dd\alpha+\sin2\chi~\sigma_3)^2\nonumber \\
&+ \cos^22\chi(1+\lambda t^4)^2(4t^2 \dd \alpha^2+(1-t^2)^2\sigma_3^2)\nonumber\\
&+(1-t^2)\big(\sin^2\chi~K_1\sigma_1^2 + \sin2\chi~ K_3\sigma_1\sigma_2 +\cos^2\chi~ K_2\sigma_2^2\big)~.
\end{align}
The axion and dilaton are given by
\begin{equation}\label{eq:axidilexpl}
\begin{split}
\rme^{\Phi} =& \frac{g_s(1+\lambda t^4)}{\sqrt{K_1 K_2-K_3^2}}\Big((1+t^2)(1-\lambda t^4)+2 t^2(1-\lambda t^2)\cos 2\chi~\cos 2\alpha\Big)\,,\\
C_0 =& -\frac{2 t^2 (1+\lambda t^2)(1-\lambda t^4)\cos 2\chi~\sin 2\alpha}{g_s(1+\lambda t^4)\big((1+t^2)(1-\lambda t^4)+2 t^2(1-\lambda t^2)\cos 2\chi~\cos 2\alpha\big)}\,.
\end{split}
\end{equation}
The NS-NS and R-R two-forms can be written compactly as
\begin{equation}
\begin{split}
B_2 + \rmi g_s C_2 =& \frac{4}{g^2} \frac{t \e^{-\rmi \alpha}}{K_1 K_2 -K_3^2}\Big[\big( a_1 \dd \chi  + a_2 \sigma_3 -\rmi\left(1- \lambda^2 t^8 \right)\left( K_1 + K_2 \right) \sin 2\chi\, \dd \alpha \big) \w \Sigma\\
&- \big( a_3 \dd \chi+a_4 \sigma_3  - \rmi \left(1- \lambda^2 t^8 \right) \left( K_1 -K_2 - 2 \rmi K_3 \right)\sin 2\chi \, \dd \alpha \big)\w \overline{\Sigma}  \Big]\,,
\end{split}
\end{equation}
where we have defined the functions
\begin{equation}
\begin{split}
a_1=&- 2 \rmi K_3 \left(1+t^2\right)\left(1- \lambda^2 t^6 \right)\,,\\
a_2 =&\,\rmi \left(1+t^2\right) \big[(K_1-K_2) \left(1-\lambda ^2t^6\right)\cos 2 \chi-2 \left(1-\lambda ^2 t^8\right)^2\\
&-2 t^2 \left(1+\lambda ^4
t^{12}-\lambda ^2t^4 \left(1+t^4\right) \right)\cos ^22 \chi  \big]\,,\\
a_3 =&\,4 t^4  \left(1-\lambda^2 t^4\right) \left(1-\lambda ^2 t^6-\lambda t^2 \left(1-t^2\right)  e^{4 \rmi \alpha }\right)\cos ^22 \chi \\
&- \left(1+t^2\right)^2 \left(1-\lambda ^2 t^8\right) \left(1-\lambda ^2 t^6+\lambda t^2  \left(1-t^2\right) e^{4 \rmi \alpha }\right)\,,\\
a_4 =&\,  \rmi \left(1-t^2\right)^2  \left(1-\lambda ^2 t^8\right) \left(1-\lambda ^2 t^6-\lambda t^2 \left(1-t^2\right)  e^{4 \rmi \alpha }\right)\cos2 \chi \,, 
\end{split}
\end{equation}
and the complex one-form $\Sigma$ is given by
\begin{align}
\Sigma = \rmi \sin \chi\, \sigma_1+ \cos \chi\, \sigma_2~.
\end{align}
The R-R four-form is
\begin{equation}
\begin{aligned}
C_4 =& \frac{32\lambda t^6 \sin4\chi(1-\lambda^2t^6 )(1-t^2)}{g_s g^4 (K_1 K_2-K_3^2)}~\sigma_1\w\sigma_2\w\Bigg[  \sin 4 \alpha~ \dd\chi \w(\sigma_3 + \sin2 \chi \dd\alpha)\\
&+\left( \frac{\left(1+\lambda  t^2\right)^2
\left(1-\lambda  t^4\right)^2}{4 \lambda  t^2 \left(1-t^2\right) \left(1-\lambda ^2 t^6\right)}-\cos^2 2 \alpha\right)\sin4 \chi~\dd \alpha \w \sigma_3\Bigg]+\frac{4}{g_sg^4}\hat{\omega}~.
\end{aligned}
\end{equation}
We have explicitly verified that all ten-dimensional equations of motion in the conventions of \cite{Bobev:2018hbq} are solved by the background above. In addition we have checked that the ten-dimensional metric in \eqref{eq:10dmetexpl}-\eqref{eq:sqspheremetexpl} and the axion and dilaton in \eqref{eq:axidilexpl} agree with the ones presented in \cite{Pilch:2000fu} which were derived using the partial uplift formulae in \cite{Pilch:2000ue}.\footnote{There is a minor misprint in Equation (6.2) of \cite{Pilch:2000fu}. The middle term on the second line of that equation should read $2a_5(v^i \dd u^i) (u^j \dd v^j)$.}

The fact that the ten-dimensional background presented above is written in terms of the one-forms $\sigma_i$ makes the $\SO(3)$ isometry of the squashed $S^5$ metric manifest. This is of course in harmony with the expectation from the dual gauge theory where in the equal mass limit of the superpotential deformation in \eqref{eq:suppotN1*} there is a manifest $\SO(3)$ flavor symmetry. We now proceed to discuss two special limits of the ten-dimensional background above in which the solution exhibits symmetry enhancement.

\subsection{Massless limit}
\label{sec:massless}

First we discuss the limit for which the mass deformation in the dual field theory vanishes, i.e. there is only a non-trivial vev for the gaugino bilinear turned on. We do not expect that this is a physically interesting limit in the gauge theory since the $\mathcal{N}=4$ SYM theory does not have a supersymmetric vacuum in which there is only a non-vanishing gaugino bilinear vev. Nevertheless it is instructive to analyze the supergravity solution above in this limit since it simplifies drastically. We note that in this limit the ten-dimensional background was also derived in Section 9 of \cite{Pilch:2000fu}. Our goal here is simply to demonstrate how this solution can be obtained from the more general solution in Section~\ref{sec:gppzuplift}.

As discussed above \eqref{eq:trange} the constant $\lambda$ can be interpreted as controlling the ratio of the gaugino bilinear vev and the mass parameter in the gauge theory. Therefore in the massless limit we should take $\lambda\to\infty$. This limit has to be taken with some care and we find that the proper procedure is to rescale $t$, $\lambda$, as well as the four-dimensional Minkowski coordinates, $x^\mu$, as
\begin{equation}
t\to \epsilon t~,\qquad \lambda \to \epsilon^{-3}\lambda~,\qquad x^\mu \to \epsilon x^\mu~,
\end{equation}
and then take $\epsilon\to0$. In this limit the functions $K_{1,2,3,4}$ defined in \eqref{eq:Kdef} simplify to
\begin{equation}
K_1\to 1~,\qquad K_2= 1~,\qquad K_3=0~,\qquad K_4= 1~.
\end{equation}
This in turn leads to the following simple form of the ten-dimensional Einstein frame metric 
\begin{equation}\label{eq:10dmetricmassless}
\begin{split}
\dd s_{10}^2 =& \frac{1}{g^2\sqrt{ g_s (1-\lambda^2t^6)}}\Big( \frac{4\dd t^2}{t^2}+\frac{1}{t^2}\big(1-\lambda^2 t^6\big)^{1/3}\dd s_4^2 \\
&\qquad\qquad\qquad+  4(1-\lambda^2 t^6)\dd s_{\mathbb{C}P^2}^2 + 4(d\alpha+\sin2\chi~\sigma_3)^2\Big)~.
\end{split}
\end{equation}
The axion-dilaton as well as the R-R four-form also take a very simple form in this limit
\begin{equation}
\rme^{\Phi}= g_s~,\qquad C_0=0~,\qquad C_4 = \frac{4}{g_sg^4}\hat{\omega}~.
\end{equation}
Finally the NS-NS and R-R two-forms read
\begin{equation}\label{eq:2frommassless}
B_2 + \rmi g_s C_2 = \frac{4 \lambda t^3}{g^2}\e^{3\rmi\alpha}~ (\dd\chi + \rmi\cos2\chi~\sigma_3)\w \overline{\Sigma}~.
\end{equation}
There is a manifest $\SU(3)$ isometry in the metric \eqref{eq:10dmetricmassless} due to the presence of the round metric on $\mathbb{C}P^2$. In addition one can check that the NS-NS and R-R fields above are also invariant under this $\SU(3)$. This is again in line with expectations from the field theory where for vanishing masses the superpotential in \eqref{eq:suppotN1*} is invariant under $\SU(3)\times \U(1)$ but the $\U(1)$ is spontaneously broken by the gaugino bilinear vev. There is however a small wrinkle in the story. The metric in \eqref{eq:10dmetricmassless} has an additional $\U(1)$ isometry generated by $\partial_{\alpha}$ this is however explicitly broken by the two-form in \eqref{eq:2frommassless}. One can however recall the type IIB supergravity is invariant under an $\SL(2,\mathbf{R})$ symmetry. This symmetry acts on the axion-dilaton $\tau=g_s(C_0+\rmi e^{-\Phi})$ and the two-forms in \eqref{eq:2frommassless} as
\begin{equation}
\left(\begin{matrix} g_sC_{2}\\ B_{2} \end{matrix}\right) \mapsto \left(\begin{matrix} a&b\\c&d \end{matrix}\right)\left(\begin{matrix} g_sC_{2}\\ B_{2} \end{matrix}\right)\,,\qquad \tau\mapsto \frac{a\tau + b}{c\tau +d}\,,\qquad ad-bc =1\,.
\end{equation}
One can then show that a shift of the coordinate $\alpha$ of the form $\alpha \to \alpha-\frac{1}{3}\delta$, where $\delta$ is a real number, combined with an $\U(1)\subset \SL(2,\mathbf{R})$ rotation of the form
\begin{equation}\label{eq:SL2matrix}
\left(\begin{matrix} \cos\delta&\sin\delta\\-\sin\delta&\cos\delta \end{matrix}\right)~,
\end{equation}
leaves all ten-dimensional fields in the massless solution above invariant. This is a supergravity manifestation of the bonus $\U(1)$ symmetry which emerges in the planar limit of the $\mathcal{N}=4$ SYM theory, see \cite{Intriligator:1998ig}.

\subsection{Vanishing gaugino vev limit}

Another limit in which the ten-dimensional background in Section~\ref{sec:gppzuplift} simplifies significantly corresponds to the limit in the gauge theory where the gaugino bilinear vev vanishes. This is achieved in the supergravity solution by simply taking the limit $\lambda\to0$ in the solution in Section~\ref{sec:gppzuplift}.

The functions $K_{1,2,3,4}$ in \eqref{eq:Kdef} take the simple form
\begin{equation}
K_1 = 1+t^2 + 2t^2\cos2\chi~,\quad K_2 = 1+t^2 - 2t^2\cos2\chi~,\quad K_3 = 0~,\quad K_4 = K_1K_2~.
\end{equation}
This means that the ten-dimensional metric in \eqref{eq:10dmetexpl}-\eqref{eq:sqspheremetexpl} simplifies to
\begin{equation}\label{eq:10dmetricvevless1}
\dd s_{10}^2 = \frac{(K_1 K_2)^{1/4}}{\sqrt{ g_s }}\left(\frac{\dd s_5^2}{(1-t^2)}+ \frac{4}{g^2 K_1 K_2}~\dd \Omega^2_5\right)~,
\end{equation}
where the squashed metrics on AdS$_5$ and $S^5$ are
\begin{equation}\label{eq:10dmetricvevless2}
\begin{split}
\dd s_5^2 &= \frac{4}{g^2t^2}\left(\dd t^2+ \left(1-t^2\right)\dd s_4^2\right)~,\\
\dd \Omega^2_5 &= K_1K_2\dd\chi^2  + 4t^2\cos^22\chi  \dd \alpha^2 + (1-t^2)\Big((\dd\alpha+\sin2\chi~\sigma_3)^2\\
&\qquad+ (1-t^2)\cos^22\chi~\sigma_3^2+\sin^2\chi~K_1\sigma_1^2 +\cos^2\chi~ K_2\sigma_2^2\Big)~.
\end{split}
\end{equation}
The axion and dilaton in this limit are given by
\begin{equation}\label{eq:axidilvevless}
\e^{\Phi} = \frac{g_s}{\sqrt{K_1K_2}}(1+t^2+2t^2 \cos 2\chi~\cos 2\alpha)~,\quad
C_0= -\frac{2t^2 \cos 2\chi~\sin 2\alpha}{g_s (1+t^2+2t^2 \cos 2\chi~\cos 2\alpha)}~,
\end{equation}
while the four-form reads
\begin{equation}
C_4= \frac{4}{g^4 g_s}\left( \frac{2t^4}{K_1K_2} \sin^2 4\chi ~\dd \alpha\w\sigma_1\w\sigma_2\w\sigma_3 +\hat{\omega}\right)~.
\end{equation}
The two-forms in this limit take the form
\begin{equation}\label{eq:2fromsvevless}
\begin{split}
&B_2 + \rmi g_s C_2 = \frac{4 t}{g^2} \e^{-\rmi\alpha}\Big(\overline{\Sigma} \w (-\dd\chi + \rmi\cos2\chi~\sigma_3) \\
& \quad +\frac{2}{K_1K_2}(-K_1\sin\chi~\sigma_1+\rmi K_2\cos\chi~\sigma_2)\w(\sin2\chi~\dd\alpha+(1-t^2\cos^22\chi)\sigma_3)\Big)\,.
\end{split}
\end{equation}
It is clear that the solution above is invariant under the $\SO(3)$ symmetry that rotates the one-forms $\sigma_i$. We also notice that the metric in \eqref{eq:10dmetricvevless1}-\eqref{eq:10dmetricvevless2} has an extra $\U(1)$ symmetry generated by $\partial_{\alpha}$. This is however broken by the explicit dependence on $\alpha$ of the axion, dilaton and the two-forms in \eqref{eq:axidilvevless}-\eqref{eq:2fromsvevless}. Similarly to the discussion at the end of Section~\ref{sec:massless} one can show that the full ten-dimensional background with $\lambda \to 0$ above is invariant under the action $\alpha \to \alpha +\delta $ combined with the $\SL(2,\mathbf{R})$ rotation in \eqref{eq:SL2matrix}.

\subsection{Zooming on the singularity}
\label{subsec:singularity}

As explained in Section \ref{sec:5d} the five-dimensional metric has a singularity as $t\to |\lambda|^{-1/3}$ for $\lambda>1$ whereas the singularity is located at $t\to 1$ for $|\lambda|\le 1$. We argued in Section \ref{sec:5d} that the five-dimensional criterion proposed in \cite{Gubser:2000nd} selects $|\lambda|\le 1$ as the physically allowed range for $\lambda$. We can provide additional evidence for this claim by looking at the $g_{tt}$ component of the Einstein frame metric \eqref{eq:10dmetexpl}. In \cite{Maldacena:2000mw} it was argued that ten-dimensional naked singularities are acceptable only when $|g_{tt}|$ is bounded as the singularity is approached. For our metric we find that  the $tt$-component of the metric for $|\lambda|>1$ blows up as
\begin{equation}
g_{tt} \sim -(|\lambda|^{-1/3}-t)^{-1/6}~,
\end{equation}
in the limit $t\to |\lambda|^{-1/3}$.

The singularity at $t=1$ for the physically acceptable range of the integration constant $|\lambda|\le 1$ is more subtle and we analyze it in some detail below. We find that the ten-dimensional metric \eqref{eq:10dmetexpl} has a singularity only at a certain locus on the squashed five-sphere, and is otherwise regular. This was already noticed in \cite{Pilch:2000fu} and can be readily observed from the ten-dimensional metric in the limit $t\to 1$:\footnote{We note that this expression for the limiting form of the metric agrees with the one in \cite{Pilch:2000fu} and differs from the one presented in \cite{Petrini:2018pjk}.  In \cite{Petrini:2018pjk} the authors performed a limiting procedure where they impose that the Ricci scalar of the limiting metric matches the limit of the \emph{full} Ricci scalar as expanded around $t \to 1$.  Our limiting procedure is defined by first diagonalizing the metric as $g = S^\top D S$ for $S \in \SO(1,9)$, and then keeping each of the eigenvalues $D$ to lowest-order around $t \to 1$ (thus by construction our limiting metric remains non-degenerate and invertible).  This procedure matches the prescription in \cite{Biquard:2015cia} for near-singularity expansions, and should be sufficient for understanding the structure of the singularity in terms of smeared brane sources.}
%
%
\begin{equation}\label{eq:ttoonemetric}
\begin{split}
\dd s_{10}^2 \approx& \frac{4\sqrt{2\sin2\chi}}{g^2\sqrt{g_s}}\left((1-\lambda^2)^{1/3} \dd s_4^2+\dd \rho^2 + \frac{\rho^2}{4}(\sigma_1^2+\sigma_2^2+\sigma_3^2)\right)\\
&+\f{4}{g^2\sqrt{g_s}(2\sin 2\chi)^{3/2}}\left(\frac{1-\lambda}{1+\lambda}\dd( \cos2\alpha~\cos 2\chi)^2 + \frac{1+\lambda}{1-\lambda}\dd(\sin2\alpha~\cos 2\chi)^2 \right)~,
\end{split}
\end{equation}
where we have introduced $\rho^2 = 2(1-t)$. Here we see that the metric is singular when we simultaneously take $\rho\to0$ (or $t\to1$) and $\chi\to0$ for any value of $|\lambda| \ne1$. We also observe that special care is required for the $t\to1$ limit when $|\lambda|=1$.

In the case of $\lambda < 1$, we can see that the singularity is actually located at $(t, \chi) = (1,0)$ rather than simply $t = 1$.  To avoid order-of-limits issues, we first change to `black ring coordinates':
\begin{equation}\label{ringcoords}
t^2 = \f{Y-X}{Y+X}~,\qquad \cos^2 2\chi = \f{Y^2-1}{Y^2-X^2}~,
\end{equation}
where the new coordinates live in the ranges $X\in(0,1)$ and $Y\in(1,\infty)$.  In these coordinates, the singularity $(t, \chi) = (1,0)$ lies at $Y \to \infty$, and the coordinate $X$ parametrizes the direction of approach to this singularity.  After expanding the metric in $1/Y$, however, one finds that the lowest-order metric can be re-organized into a double limit in the original $(t,\chi)$ coordinates, which we write (again in terms of $\rho^2 = 2(1-t)$) as
\begin{equation}\label{eq:lambdalessthanonemetric}
\dd s_{10}^2 \approx \frac{4\Delta\rho^2}{g^2\sqrt{g_s}}\bigg((1-\lambda^2)^{1/3}\dd s_4^2  + \rho^{-2}\dd \alpha^2+\dd \rho^2 + \f{\rho^2}{4} (\sigma_2^2+\sigma_3^2) 
 + \f{(1-\lambda)^2(\dd\chi^2+\chi^2\sigma_1^2)}{1+\lambda^2+2\lambda\cos 4\alpha}\bigg)~,
\end{equation}
where
\begin{equation}
\Delta^4 =\frac{4(1+\lambda^2+2\lambda\cos 4\alpha)}{(1-\lambda^2)\rho^{6}}~.
\end{equation}
It was suggested in \cite{Pilch:2000fu} that the metric has the expected local form of seven-branes on a ring parametrised by $\alpha$.  This conclusion was reached by observing that the metric in \eqref{eq:ttoonemetric} prefers to be written in $8+2$ form as
\begin{equation}\label{lambdalessthanonemetricsplit}
h^{-1/4}\dd s^2_{\mathbb{R}^{1,3}\times  \mathbb{R}^4/\mathbb{Z}_2} + h^{3/4} \dd s^2_{D^2}~,
\end{equation}
where $D^2$ denotes the disc spanned by the coordinates $\chi$ and $\alpha$ and the discrete modding is due to the fact that the one-forms $\sigma_i$ only span $\SO(3)\simeq S^3/\mathbb{Z}_2$ and not $\SU(2)\simeq S^3$.  However, while this metric exhibits an $8+2$ splitting, the function $h=(2\sin 2\chi)^{-2}$ appears instead with powers that resemble the `harmonic function' for a metric describing five-branes.  The structure of the near-singularity metric does not seem very elucidating.  By looking at the other type IIB supergravity fields in the limit $(t,\chi)\to(1,0)$ we see that the axion and dilaton diverge according to
\begin{equation}
\e^{\Phi} \approx \frac{4g_s \cos^2\alpha}{\Delta^2\rho^4}~,\quad
C_0 \approx -g_s^{-1}\tan\alpha~.
\end{equation}
The axion-dilaton can diverge for either five-branes or seven-branes; seven-branes, however, would have a logarithmic divergence in order to produce non-trivial monodromies in the axion-dilaton matrix integrated around the singularity.  We also see that the other form fields
\begin{equation}
B_2 + \rmi g_s C_2 \approx \frac{4 \rmi\e^{-\rmi\alpha}}{g^2}\sigma_2\w\left(\rmi\dd\chi+\sigma_3\right)~,\quad
C_4 \approx -\frac{4}{ g^{4} g_s}\dd\alpha\w\sigma_1\w\sigma_2\w\sigma_3~,
\end{equation}
are regular in this limit. Note however that the energy density of $H_3 = \dd B_2$ does diverge.  In summary, it would seem reasonable to give this limit the interpretation of the near-horizon geometry of smeared D5/NS5-branes. This interpretation must be supported by matching the rate at which various fields diverge as $\rho\to0$ and by demonstrating that the charge of the smeared branes is correctly reproduced. We will revisit this analysis in a future publication \cite{INPROG}.

The ten-dimensional solution presented above does not simplify in any obvious way when taking $\lambda \to 1$, nevertheless two notable features are important in this limit. First of all we remember that $\lambda \to 1$ is at the edge of the physically acceptable range for $\lambda$ as discussed above. It has therefore been suggested in \cite{Pilch:2000fu} and \cite{Gubser:2000nd} that the solution for $\lambda=1$ plays a key role for the dual field theory. From the ten-dimensional perspective we see that the singularity is not ringlike as we saw for $\lambda<1$ but rather there are two degrees of singularity in the metric. Just as for the five-dimensional metric the ten-dimensional metric exhibits a singularity as $t\to1$. This is independent of the location on the five-sphere as $t$ approaches $1$. The metric has a greater degree of singularity located at $(t,\chi,\alpha) = (1,0,\pm \pi/4)$. For the analysis at hand we would have to carefully parametrize the approach towards the singularity using coordinates similar to the ones in \eqref{ringcoords}. In our short analysis presented here we choose a particular approach towards the singularity similar to the one in \cite{Pilch:2000fu}. To this end we let $\alpha = \pi/4 + \psi^2$ and we take $\psi \sim \chi \sim (1-t)\sim\epsilon \to0$. In this limit we find that the dilaton blows up while the axion is regular,
\begin{equation}
\e^{\Phi}\sim\frac{1}{\epsilon}~,\qquad C_0 \sim \f{1}{g_s}~.
\end{equation}
From this alone we see that the singularity can not have the interpretation of D7-branes since the axion never diverges. This is different from the behaviour for $\lambda<1$ where for $\alpha\to \pi/2$ the axion diverges. Furthermore the energy density for the NS-NS three-form is regular in this limit
\begin{equation}
\e^{-\Phi} |H_3|^2 \sim \e^{\Phi} |F_3|^2\sim\epsilon^0~.
\end{equation}
The supergravity solutions for general $(p,q)$ fivebranes in flat space have singular energy densities as the near-brane region is approached. Thus it is not clear whether the $\lambda=1$ solution can be interpreted as the holographic dual of one of the $\mathcal{N}=1^*$ vacua in the description of Polchinski and Strassler in \cite{Polchinski:2000uf}.

\section{Discussion}
\label{sec:discussion}

In this work we exploited a consistent truncation of the five-dimensional $\mathcal{N}=8$ $\SO(6)$ gauged supergravity as well as the explicit results in \cite{Hohm:2013vpa,Baguet:2015sma,Baguet:2015xha} to uplift the well-known GPPZ solution of the five-dimensional theory \cite{Girardello:1999bd} to a full ten-dimensional background of type IIB supergravity. This provides a rare example of an explicit analytic solution of type IIB supergravity which is holographically dual to a non-conformal deformation of the $\mathcal{N}=4$ SYM theory in the planar limit. There are several interesting directions for further research which we summarize briefly below.

As expected from the original five-dimensional construction in \cite{Girardello:1999bd} our ten-dimensional solution preserves an $\SO(3)$ symmetry realized as an isometry of the squashed $S^5$ internal space. This is in harmony with the dual field theory where we have a mass deformation of the form \eqref{eq:suppotN1*} with all three masses being equal. Clearly there are more general $\mathcal{N}=1^{*}$ deformations that can be explored. When $m_3=0$ and $m_1=m_2$  supersymmetry is enhanced and the theory is known as $\mathcal{N}=2^*$. Its ten-dimensional supergravity dual was constructed explicitly in \cite{Pilch:2000ue}. Another special limit of the deformation in \eqref{eq:suppotN1*} is provided by taking $m_2=m_3=0$ and general $m_1$. This theory preserves $\mathcal{N}=1$ supersymmetry and as found in \cite{Leigh:1995ep} flows to an interacting SCFT in the IR. The supergravity dual of this RG flow, which interpolates between two supersymmetic AdS$_5$ vacua, was first studied in five dimensions in \cite{Freedman:1999gp} and then uplifted to a solution of type IIB supergravity in \cite{Pilch:2000fu}. For general non-vanishing values of the masses it is harder to construct explicit supergravity solutions but some progress was made in \cite{Khavaev:2000gb} where several numerical solutions of the five-dimensional $\mathcal{N}=8$ $\SO(6)$ gauged supergravity were constructed. It will be interesting to revisit this question and construct the most general five-dimensional solution dual to $\mathcal{N}=1^{*}$ with general values for the mass parameters and arbitrary vevs compatible with supersymmetry. The truncation with 18 scalar fields discussed in \cite{Bobev:2016nua} should contain many of these interesting solutions. Once the five-dimensional solutions are constructed it should be possible to uplift them to ten-dimensional supergravity following the approach we utilized in this work. However it should be expected that the ten-dimensional metric and matter fields take an unwieldy form due to the lack of any continuous global symmetries in the dual gauge theory. A more accessible goal should be to find the type IIB supergravity uplift of the five-dimensional solutions in \cite{Bobev:2016nua} which are holographically dual to the $\mathcal{N}=1^{*}$ SYM theory placed on the round $S^4$. In particular one of the solutions in  \cite{Bobev:2016nua}, dual to the equal mass $\mathcal{N}=1^{*}$ theory on $S^4$, is contained within the four-scalar supergravity truncation described in Section~\ref{subsec:gppz}. Unlike the GPPZ solution this five-dimensional solution has non-trivial profiles for all four scalars in the truncation and is only numerically known. Nevertheless it is invariant under the same $\SO(3)$ symmetry that the GPPZ solution preserves thus we expect that the ten-dimensional uplift should be within reach.

While the ten-dimensional metric, the axion, and the dilaton of the solution presented in Section~\ref{sec:gppzuplift} were already derived in \cite{Pilch:2000fu} we have now found explicitly the NS-NS and R-R forms of the full type IIB background. This should be particularly useful if one is interested in studying the physics of this solution using probe strings and branes. This is especially relevant for understand the singularity of the solution in the IR region and possible mechanisms for its resolution. There are at least two clear physical methods to remove the singularity of the ten-dimensional solution. One possibility is to eschew supersymmetry and introduce an IR cutoff in the gauge theory by turning on finite temperature. In the dual supergravity this should correspond to finding a black hole background which is asymptotic to the supersymmetric solution in Section~\ref{sec:gppzuplift} in the UV region and has a regular horizon which shields the naked singularity in the IR. Some progress in constructing such a black brane solution was reported in \cite{Freedman:2000xb} but the fully backreacted finite temperature background is still not known. Alternatively, as pointed out in \cite{Bobev:2013cja,Bobev:2016nua} and emphasized in \cite{Bobev:2018ugk}, one can introduce an IR cutoff of the gauge theory which preserves four real supercharges by placing it on $S^4$. In the ten-dimensional supergravity this is realized by a smooth Euclidean solution in which the naked singularity of the solution in Section~\ref{sec:gppzuplift} is replaced by a smooth cap-off in the IR region.

Perhaps the most interesting open question is to find how the type IIB background in Section~\ref{sec:gppzuplift} relates to the work of Polchinski-Strassler \cite{Polchinski:2000uf}. It was argued in \cite{Polchinski:2000uf} that many of the supersymmetric vacua  of the planar $\mathcal{N}=1^{*}$ theory are captured by asymptotically AdS$_5\times S^5$ type IIB string theory/supergravity solutions which have D3-branes polarized into 5-branes in the IR region. The authors of \cite{Polchinski:2000uf} present many beautiful arguments for this picture but unfortunately there are no known explicit supergravity solutions that demonstrate it rigorously. In similar non-conformal holographic models it was  found that supergravity solutions which exhibit naked singularities can be replaced by fully regular supersymmetric solutions with the same UV asymptotics. Two well-known examples of such solutions dual to four-dimensional $\mathcal{N}=1$ gauge theories are the KS solution \cite{Klebanov:2000hb}, which resolves the singular KT solution \cite{Klebanov:2000nc}, and the solution in \cite{Maldacena:2000yy} dual to a vacuum of the pure $\mathcal{N}=1$ SYM theory. Perhaps the closest analog to the singular GPPZ solution, for which there is an explicitly known resolution mechanism, is the singular solution in \cite{Pope:2003jp} which is dual to a supersymmetric mass deformation of the ABJM theory. The resolution of this solution into a smooth background of eleven-dimensional supergravity with non-trivial topology was found in \cite{Bena:2004jw,Lin:2004nb}. It will be most interesting to exhibit a similar supergravity solution that generalizes the GPPZ solution and provides an explicit realization of the polarization mechanism envisioned in \cite{Polchinski:2000uf}.

\bigskip
\bigskip
\leftline{\bf Acknowledgements}
\smallskip
\noindent We would like to thank Iosif Bena, Henriette Elvang, Prem Kumar, Krzysztof Pilch, Silviu Pufu, and Jorge Santos for useful discussions. The work of NB is supported in part by an Odysseus grant G0F9516N from the FWO. FFG is supported by a postdoctoral fellowship from the Fund for Scientific Research - Flanders (FWO). The work of BEN has been supported variously by ERC grant ERC-2011-StG 279363-HiDGR, and by ERC grant ERC-2013-CoG 616732-HoloQosmos, and by the FWO and European Union's Horizon 2020 research and innovation program under the Marie Sk\l{}odowska-Curie grant agreement No. 665501. BEN is an FWO [PEGASUS]$^2$ Marie Sk\l{}odowska-Curie Fellow. The work of JvM is supported by a doctoral fellowship from the Fund for Scientific Research - Flanders (FWO). We are also supported by the KU Lueven C1 grant ZKD1118 C16/16/005.

\bibliography{N1starIIB}

\providecommand{\href}[2]{#2}\begingroup\raggedright\begin{thebibliography}{10}

\bibitem{Girardello:1998pd}
L.~Girardello, M.~Petrini, M.~Porrati, and A.~Zaffaroni, {\it {Novel local CFT
  and exact results on perturbations of N=4 superYang Mills from AdS
  dynamics}},  {\em JHEP} {\bf 12} (1998) 022,
  [\href{http://arxiv.org/abs/hep-th/9810126}{{\tt hep-th/9810126}}].

\bibitem{Freedman:1999gp}
D.~Z. Freedman, S.~S. Gubser, K.~Pilch, and N.~P. Warner, {\it {Renormalization
  group flows from holography supersymmetry and a c-theorem}},  {\em Adv.
  Theor. Math. Phys.} {\bf 3} (1999) 363--417,
  [\href{http://arxiv.org/abs/hep-th/9904017}{{\tt hep-th/9904017}}].

\bibitem{Kim:1985ez}
H.~J. Kim, L.~J. Romans, and P.~van Nieuwenhuizen, {\it {The mass spectrum of
  chiral $N=2$ $D=10$ supergravity on $S^5$}},  {\em Phys. Rev.} {\bf D32}
  (1985) 389.

\bibitem{Gunaydin:1984qu}
M.~Gunaydin, L.~J. Romans, and N.~P. Warner, {\it {Gauged N=8 Supergravity in
  Five-Dimensions}},  {\em Phys. Lett.} {\bf 154B} (1985) 268--274.

\bibitem{Gunaydin:1985cu}
M.~Gunaydin, L.~J. Romans, and N.~P. Warner, {\it {Compact and Noncompact
  Gauged Supergravity Theories in Five-Dimensions}},  {\em Nucl. Phys.} {\bf
  B272} (1986) 598--646.

\bibitem{Pernici:1985ju}
M.~Pernici, K.~Pilch, and P.~van Nieuwenhuizen, {\it {Gauged N=8 D=5
  Supergravity}},  {\em Nucl. Phys.} {\bf B259} (1985) 460.

\bibitem{Cvetic:2000nc}
M.~Cveti\v{c}, H.~L{\"u}, C.~N. Pope, A.~Sadrzadeh, and T.~A. Tran, {\it
  {Consistent SO(6) reduction of type IIB supergravity on $S^5$}},  {\em Nucl.
  Phys.} {\bf B586} (2000) 275--286,
  [\href{http://arxiv.org/abs/hep-th/0003103}{{\tt hep-th/0003103}}].

\bibitem{Pilch:2000ue}
K.~Pilch and N.~P. Warner, {\it {N=2 supersymmetric RG flows and the IIB
  dilaton}},  {\em Nucl. Phys.} {\bf B594} (2001) 209--228,
  [\href{http://arxiv.org/abs/hep-th/0004063}{{\tt hep-th/0004063}}].

\bibitem{Lee:2014mla}
K.~Lee, C.~Strickland-Constable, and D.~Waldram, {\it {Spheres, generalised
  parallelisability and consistent truncations}},  {\em Fortsch. Phys.} {\bf
  65} (2017), no.~10-11 1700048, [\href{http://arxiv.org/abs/1401.3360}{{\tt
  arXiv:1401.3360}}].

\bibitem{Baguet:2015sma}
A.~Baguet, O.~Hohm, and H.~Samtleben, {\it {Consistent Type IIB Reductions to
  Maximal 5D Supergravity}},  {\em Phys. Rev.} {\bf D92} (2015), no.~6 065004,
  [\href{http://arxiv.org/abs/1506.01385}{{\tt arXiv:1506.01385}}].

\bibitem{Hohm:2013vpa}
O.~Hohm and H.~Samtleben, {\it {Exceptional Field Theory I: $E_{6(6)}$
  covariant Form of M-Theory and Type IIB}},  {\em Phys. Rev.} {\bf D89}
  (2014), no.~6 066016, [\href{http://arxiv.org/abs/1312.0614}{{\tt
  arXiv:1312.0614}}].

\bibitem{Baguet:2015xha}
A.~Baguet, O.~Hohm, and H.~Samtleben, {\it {E$_{6(6)}$ Exceptional Field
  Theory: Review and Embedding of Type IIB}},  {\em PoS} {\bf CORFU2014} (2015)
  133, [\href{http://arxiv.org/abs/1506.01065}{{\tt arXiv:1506.01065}}].

\bibitem{Bobev:2018hbq}
N.~Bobev, F.~F. Gautason, and J.~Van~Muiden, {\it {Precision Holography for
  $\mathcal{N}=2^{*}$ on $S^4$ from type IIB Supergravity}},  {\em JHEP} {\bf
  04} (2018) 148, [\href{http://arxiv.org/abs/1802.09539}{{\tt
  arXiv:1802.09539}}].

\bibitem{Bobev:2013cja}
N.~Bobev, H.~Elvang, D.~Z. Freedman, and S.~S. Pufu, {\it {Holography for
  $\mathcal{N} = 2^*$ on $S^4$}},  {\em JHEP} {\bf 07} (2014) 001,
  [\href{http://arxiv.org/abs/1311.1508}{{\tt arXiv:1311.1508}}].

\bibitem{Girardello:1999bd}
L.~Girardello, M.~Petrini, M.~Porrati, and A.~Zaffaroni, {\it {The Supergravity
  dual of N=1 superYang-Mills theory}},  {\em Nucl. Phys.} {\bf B569} (2000)
  451--469, [\href{http://arxiv.org/abs/hep-th/9909047}{{\tt hep-th/9909047}}].

\bibitem{Bobev:2016nua}
N.~Bobev, H.~Elvang, U.~Kol, T.~Olson, and S.~S. Pufu, {\it {Holography for $
  \mathcal{N} $ = 1$^{*}$ on S$^{4}$}},  {\em JHEP} {\bf 10} (2016) 095,
  [\href{http://arxiv.org/abs/1605.00656}{{\tt arXiv:1605.00656}}].

\bibitem{Pilch:2000fu}
K.~Pilch and N.~P. Warner, {\it {N=1 supersymmetric renormalization group flows
  from IIB supergravity}},  {\em Adv. Theor. Math. Phys.} {\bf 4} (2002)
  627--677, [\href{http://arxiv.org/abs/hep-th/0006066}{{\tt hep-th/0006066}}].

\bibitem{Polchinski:2000uf}
J.~Polchinski and M.~J. Strassler, {\it {The String dual of a confining
  four-dimensional gauge theory}},
  \href{http://arxiv.org/abs/hep-th/0003136}{{\tt hep-th/0003136}}.

\bibitem{Myers:1999ps}
R.~C. Myers, {\it {Dielectric branes}},  {\em JHEP} {\bf 12} (1999) 022,
  [\href{http://arxiv.org/abs/hep-th/9910053}{{\tt hep-th/9910053}}].

\bibitem{Dorey:2000fc}
N.~Dorey and S.~P. Kumar, {\it {Softly broken N=4 supersymmetry in the large N
  limit}},  {\em JHEP} {\bf 02} (2000) 006,
  [\href{http://arxiv.org/abs/hep-th/0001103}{{\tt hep-th/0001103}}].

\bibitem{Dorey:1999sj}
N.~Dorey, {\it {An Elliptic superpotential for softly broken N=4 supersymmetric
  Yang-Mills theory}},  {\em JHEP} {\bf 07} (1999) 021,
  [\href{http://arxiv.org/abs/hep-th/9906011}{{\tt hep-th/9906011}}].

\bibitem{Aharony:2000nt}
O.~Aharony, N.~Dorey, and S.~P. Kumar, {\it {New modular invariance in the N=1*
  theory, operator mixings and supergravity singularities}},  {\em JHEP} {\bf
  06} (2000) 026, [\href{http://arxiv.org/abs/hep-th/0006008}{{\tt
  hep-th/0006008}}].

\bibitem{Gubser:2000nd}
S.~S. Gubser, {\it {Curvature singularities: The Good, the bad, and the
  naked}},  {\em Adv. Theor. Math. Phys.} {\bf 4} (2000) 679--745,
  [\href{http://arxiv.org/abs/hep-th/0002160}{{\tt hep-th/0002160}}].

\bibitem{Maldacena:2000mw}
J.~M. Maldacena and C.~Nunez, {\it {Supergravity description of field theories
  on curved manifolds and a no go theorem}},  {\em Int. J. Mod. Phys.} {\bf
  A16} (2001) 822--855, [\href{http://arxiv.org/abs/hep-th/0007018}{{\tt
  hep-th/0007018}}]. [,182(2000)].

\bibitem{Petrini:2018pjk}
M.~Petrini, H.~Samtleben, S.~Schmidt, and K.~Skenderis, {\it {The 10d Uplift of
  the GPPZ Solution}},  \href{http://arxiv.org/abs/1805.01919}{{\tt
  arXiv:1805.01919}}.

\bibitem{Leigh:1995ep}
R.~G. Leigh and M.~J. Strassler, {\it {Exactly marginal operators and duality
  in four-dimensional $N=1$ supersymmetric gauge theory}},  {\em Nucl. Phys.}
  {\bf B447} (1995) 95--136, [\href{http://arxiv.org/abs/hep-th/9503121}{{\tt
  hep-th/9503121}}].

\bibitem{Warner:1983vz}
N.~P. Warner, {\it {Some New Extrema of the Scalar Potential of Gauged $N=8$
  Supergravity}},  {\em Phys. Lett.} {\bf 128B} (1983) 169--173.

\bibitem{Khavaev:1998fb}
A.~Khavaev, K.~Pilch, and N.~P. Warner, {\it {New vacua of gauged N=8
  supergravity in five-dimensions}},  {\em Phys. Lett.} {\bf B487} (2000)
  14--21, [\href{http://arxiv.org/abs/hep-th/9812035}{{\tt hep-th/9812035}}].

\bibitem{KP}
K.~Pilch, {\it {unpublished notes}}, .

\bibitem{Intriligator:1998ig}
K.~A. Intriligator, {\it {Bonus symmetries of N=4 superYang-Mills correlation
  functions via AdS duality}},  {\em Nucl. Phys.} {\bf B551} (1999) 575--600,
  [\href{http://arxiv.org/abs/hep-th/9811047}{{\tt hep-th/9811047}}].

\bibitem{Biquard:2015cia}
O.~Biquard, {\it {M\'etriques hyperk\"ahl\'eriennes pli\'ees}},
  \href{http://arxiv.org/abs/1503.04128}{{\tt arXiv:1503.04128}}.

\bibitem{INPROG}
N.~Bobev, F.~F. Gautason, B.~E. Niehoff, and J.~van Muiden, {\it {In
  progress}}, .

\bibitem{Khavaev:2000gb}
A.~Khavaev and N.~P. Warner, {\it {A Class of N=1 supersymmetric RG flows from
  five-dimensional N=8 supergravity}},  {\em Phys. Lett.} {\bf B495} (2000)
  215--222, [\href{http://arxiv.org/abs/hep-th/0009159}{{\tt hep-th/0009159}}].

\bibitem{Freedman:2000xb}
D.~Z. Freedman and J.~A. Minahan, {\it {Finite temperature effects in the
  supergravity dual of the N=1* gauge theory}},  {\em JHEP} {\bf 01} (2001)
  036, [\href{http://arxiv.org/abs/hep-th/0007250}{{\tt hep-th/0007250}}].

\bibitem{Bobev:2018ugk}
N.~Bobev, P.~Bomans, and F.~F. Gautason, {\it {Spherical Branes}},
  \href{http://arxiv.org/abs/1805.05338}{{\tt arXiv:1805.05338}}.

\bibitem{Klebanov:2000hb}
I.~R. Klebanov and M.~J. Strassler, {\it {Supergravity and a confining gauge
  theory: Duality cascades and chi SB resolution of naked singularities}},
  {\em JHEP} {\bf 08} (2000) 052,
  [\href{http://arxiv.org/abs/hep-th/0007191}{{\tt hep-th/0007191}}].

\bibitem{Klebanov:2000nc}
I.~R. Klebanov and A.~A. Tseytlin, {\it {Gravity duals of supersymmetric SU(N)
  x SU(N+M) gauge theories}},  {\em Nucl. Phys.} {\bf B578} (2000) 123--138,
  [\href{http://arxiv.org/abs/hep-th/0002159}{{\tt hep-th/0002159}}].

\bibitem{Maldacena:2000yy}
J.~M. Maldacena and C.~Nunez, {\it {Towards the large N limit of pure N=1
  superYang-Mills}},  {\em Phys. Rev. Lett.} {\bf 86} (2001) 588--591,
  [\href{http://arxiv.org/abs/hep-th/0008001}{{\tt hep-th/0008001}}].

\bibitem{Pope:2003jp}
C.~N. Pope and N.~P. Warner, {\it {A Dielectric flow solution with maximal
  supersymmetry}},  {\em JHEP} {\bf 04} (2004) 011,
  [\href{http://arxiv.org/abs/hep-th/0304132}{{\tt hep-th/0304132}}].

\bibitem{Bena:2004jw}
I.~Bena and N.~P. Warner, {\it {A Harmonic family of dielectric flow solutions
  with maximal supersymmetry}},  {\em JHEP} {\bf 12} (2004) 021,
  [\href{http://arxiv.org/abs/hep-th/0406145}{{\tt hep-th/0406145}}].

\bibitem{Lin:2004nb}
H.~Lin, O.~Lunin, and J.~M. Maldacena, {\it {Bubbling AdS space and 1/2 BPS
  geometries}},  {\em JHEP} {\bf 10} (2004) 025,
  [\href{http://arxiv.org/abs/hep-th/0409174}{{\tt hep-th/0409174}}].

\end{thebibliography}\endgroup
\bibliographystyle{JHEP}

\end{document}